\documentclass[superscriptaddress,
 reprint,
 amsmath,amssymb,
 aps,showpacs
]{revtex4-1}

\usepackage{graphicx}
\usepackage{dcolumn}
\usepackage{bm}
\usepackage{hyperref}
\usepackage{amsmath,amssymb}

\usepackage{comment}
\usepackage{xcolor}

\begin{document}

\preprint{APS/123-QED}

\title{Prediction of excitable wave dynamics using machine learning}

\author{Mahesh Kumar Mulimani}
\affiliation{Department of Physics, University of California San Diego, La Jolla, CA, United States } 
\author{Sebastian Echeverria-Alar}
\affiliation{Department of Physics, University of California San Diego, La Jolla, CA, United States } 
\author{Michael Reiss}
\affiliation{Department of Bioengineering, University of California San Diego, La Jolla, CA, United States } 
\author{Wouter-Jan Rappel }
 \affiliation{Department of Physics, University of California San Diego, La Jolla, CA, United States } 

\date{\today}
             
\begin{abstract}
Excitable systems can exhibit a variety of dynamics with different complexity, ranging from a single, stable spiral to spiral defect chaos (SDC), during which 
spiral waves are continuously formed and destroyed. 
The corresponding reaction-diffusion models, including ones for cardiac tissue,
can involve a large number of variables and  can be time-consuming to simulate. 
Here we trained a deep-learning (DL) model using snapshots from a single variable, obtained by simulating 
a single quasi-periodic spiral wave and SDC using a generic
cardiac model. 
Using the trained DL model, we predicted the dynamics in both cases, using timesteps that are much 
larger than required for the simulations of the underlying equations. 
We show that the DL model is able to predict the trajectory of a quasi-periodic spiral wave and that 
the SDC activaton patterns can be predicted for approximately one Lyapunov time. 
Furthermore, we show that the DL model accurately captures the statistics of termination events in 
SDC, including the mean termination time. Finally, we show that a DL model trained using a specific domain size is able to replicate termination statistics on larger domains, resulting in significant computational savings. 
\end{abstract}

\maketitle

Spiral waves are inherent dynamical solutions in spatially extended excitable systems and are commonly observed in various physical and chemical systems, including Rayleigh-Bénard convection cells \cite{morris1996spatio}, the Belousov-Zhabotinsky reaction \cite{qiao2009defect}, and catalytic reaction systems \cite{beta2006defect}, and in prototypical nonlinear equations like the complex Ginzburg-Landau equation \cite{huepe2004statistics}.
These waves can occur in the form of periodic solutions, during which the spiral wave rotates with a 
fixed frequency, or as quasi-periodic (QP) solutions, during which multiple frequencies govern the dynamics of 
the waves. In addition, spiral waves can become unstable, resulting in spiral defect chaos (SDC).
During SDC, spiral waves continuously form through wave break and are annihilated upon collision with other spiral waves or non-conducting boundaries. These stochastic creation and annihilation events sustain SDC until the last spiral wave disappears.

Spiral waves can also be present in cardiac tissue, where they 
have been shown to be a crucial component in the maintenance of severe cardiac arrhythmias \cite{davidenko1992stationary,nattel2017demystifying,christoph2018electromechanical,Naretal12b,karma2013physics,rappel2022thephysics}. Single spiral waves underlie tachycardia, abnormally fast heart rhythms, 
while SDC may be responsible for fibrillation, during which the heart is unable to pump coherently.
In SDC, the competition between the formation and annihilation of spiral waves due to the tissue's excitable nature critically determines the mean termination time, $\tau$ \cite{vidmar2019extinction,dharmaprani2019renewal,lilienkamp2020terminating}. This termination time  is a stochastic quantity and 
computational studies have shown that it is exponentially distributed
\cite{vidmar2019extinction}.
The mean termination time can be computed using statistical physics 
techniques that solve the birth-death equation that describes the evolution of the 
number of spiral waves in the system \cite{vidmar2019extinction}. 
Consistent with numerical results \cite{qu2006critical}, this analysis revealed that $\tau$ increases exponentially with the area of the tissue.

\begin{figure}
\includegraphics[scale=0.135]{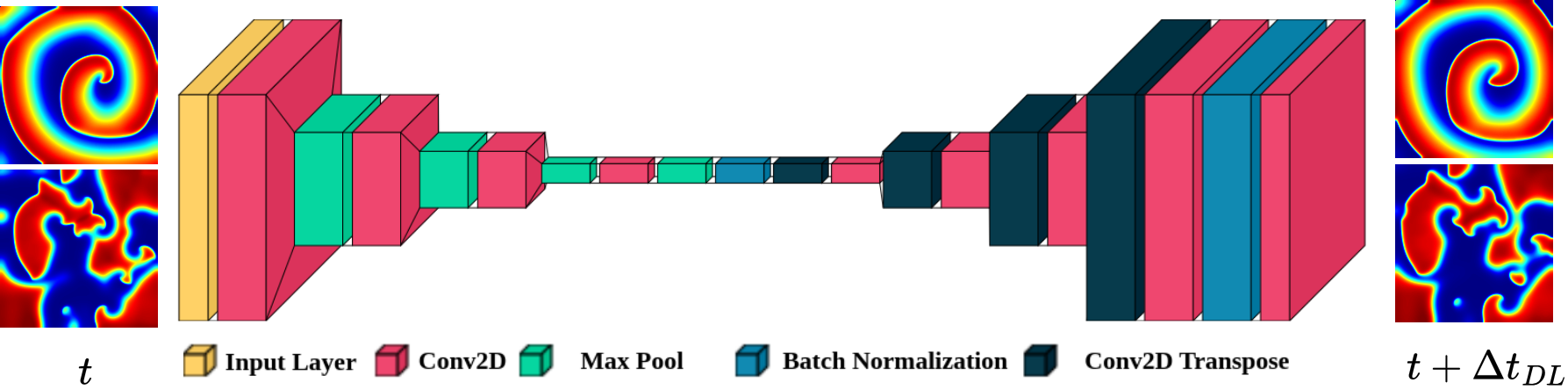}
\caption{\label{fig:fig1} Schematic figure illustrating the deep learning Encoder-Decoder model with the two different wave patterns used to train the model. The input is a snapshot of either a stable spiral wave (upper panel) or of SDC (lower panel) at time $t$, while the output is the snapshot corresponding to the patterns at $t+\Delta t_{DL}$. In this study, we used $\Delta t_{DL}=5, 10$,  \ \text{and} $15 \ ms$.}
\label{fig:encoder}
\end{figure}

A large number of mathematical models have been developed to study the dynamics of spiral waves in excitable 
systems in general, and in cardiac tissue in particular~\cite{ten2006alternans,o2011simulation,asakura2014ead}. 
These models typically consist of many variables and stiff equations, requiring small time steps. 
In this study, we show that a deep learning (DL) approach trained on snapshots of a single 
variable can be used to predict the dynamics of complex 
activation patterns. Specifically, it is able to predict 
QP spiral waves and the spatio-temporal patterns during SDC  using time steps
that are many orders of magnitude larger than the time step required for the PDE model.
Using only one variable for training is important for applications to clinical or experimental data, which generally only measure a 
single quantity, while the employing 
large timesteps can result in a decrease in computational time.
Importantly, our DL approach is independent of the underlying model  and we  illustrate it here using a generic cardiac electrophysiological model.

We trained a convolution DL encoder-decoder model, schematically shown in Fig. \ref{fig:encoder}, using data from simulations of a 
cardiac model. 
This is a type of neural network architecture that combines convolutional layers for feature extraction and transformation with an encoder-decoder framework for learning and generating complex mappings between input and output data \cite{lecun2015deep}. It
excels in tasks where the input and output data share spatial or temporal dependencies, such as in medical image analysis, semantic segmentation, and natural language processing, and finds 
 applications in various tasks, including image segmentation, image-to-image translation, generative modeling, and sequence-to-sequence prediction \cite{mayer2016large,yang2016object,badrinarayanan2017segnet}. 
For the electrophysiological model, we chose 
 the Fenton-Karma (FK) model, a widely used cardiac model \cite{fenton1998vortex}.
The FK model consists of three variables, including the variable $u$ that represents the membrane potential. The membrane potential obeys the 
partial differential equation (PDE) 
\begin{eqnarray}
	\frac{\partial{u}}{\partial{t}} = D {\nabla^2}u - \frac{I_{ion}}{C_{m}}
	\label{eqn:pde}
\end{eqnarray}
where $C_m$ ($\mu \rm{F}\, cm^{-2}$) is the membrane capacitance,  and the diffusive term  expresses the inter-cellular
coupling via gap junctions and diffusion constant $D$. 
The reaction term in the PDE describes the
membrane currents $I_{ion}$ of the model. These currents depend on $u$ through 
nonlinear equations. Details of the model, together with the 
two sets of parameter values used in our simulations, 
are presented in the Supplemental Material (SM) and Table S1 \cite{SupMat_SDC}.

To generate data for the training of the DL model, we simulated Eqn. \ref{eqn:pde} using a domain with no-flux boundary and of the size of $N\times N$ with  $N=180$. We employed a five-point stencil algorithm for the Laplacian and a forward Euler method for the advancement of time.
We chose a spatial discretization of $\Delta x=0.025 cm$ and 
a temporal  discretization
$\Delta t_{PDE} = 0.1 \ ms$, which was close to the von Neumann stability 
criterion for the value of $D$ we considered ($D=0.001$ cm$^2$/s).
We used two different parameter sets: one that resulted in a QP spiral wave with multiple frequencies and one for which spiral waves exhibited break-up, corresponding to SDC. 
To generate training set data for the QP spiral wave, we simulated $n=50$ 
sequences of length 8s in which the location of the spiral wave was started at a randomly chosen position.
For the SDC simulations, we generated training data by 
starting with a dynamical state exhibiting multiple spiral waves and creating $n=500$  independent initial conditions by adding uniformly distributed noise with an amplitude $A_{noise}=10^{-5}$ to all three variables and at every 
 grid point. These initial conditions were then evolved for $1 s$ to remove transients, 
 after the simulations continued until SDC terminated. 

In contrast to an earlier study that used all variables of a cardiac PDE model to train a neural network  \cite{zimmermann2018observing}, our DL used only a single variable and its snapshots as an input. 
We chose here $u$, motivated
by the fact that  in experiments and clinical recordings, only one variable, typically the membrane potential, can be measured. 
In both the QP and SDC case, the DL model was trained on 400 successive snapshots $\Delta t_{DL}$ 
apart, taken from 50 randomly selected simulations. We tested several values of 
$\Delta t_{DL}$, all   taken to be much larger than the time step required to 
solve the underlying stiff PDEs. 
It is important to emphasize here that  the training dataset did not contain a termination event.
The dataset was split into training and validation datasets in a  $70\%$ and $30\%$ proportion. Thus, the training dataset consisted of snapshots from 35 simulations and the validation dataset consisted of 
snapshots of 15 simulations. Testing of the DL model was performed using the 
remaining 450 simulations. 
To verify that the results from our DL model are consistent, we performed a five-fold cross-validation check, where the training dataset for each session was different and were randomly picked out of 500 SDC simulations.
Further  training details, such as hyperparameters, choice of the optimization algorithm, and the loss function specifics, are detailed in the Supplemental Material and Fig. S1~\cite{SupMat_SDC}.

Once the DL model was trained, we employed it to perform predictions of the wave patterns. For this, we started with an initial snapshot as an input to the DL model and predicted the next snapshot after $\Delta t_{DL}$. This predicted snapshot was then used to predict the next snapshot, again $\Delta t_{DL}$  later. This process was 
repeated for a fixed amount of time for the QP spiral and until termination was achieved for  SDC.

\begin{figure}
\includegraphics[scale=0.270]{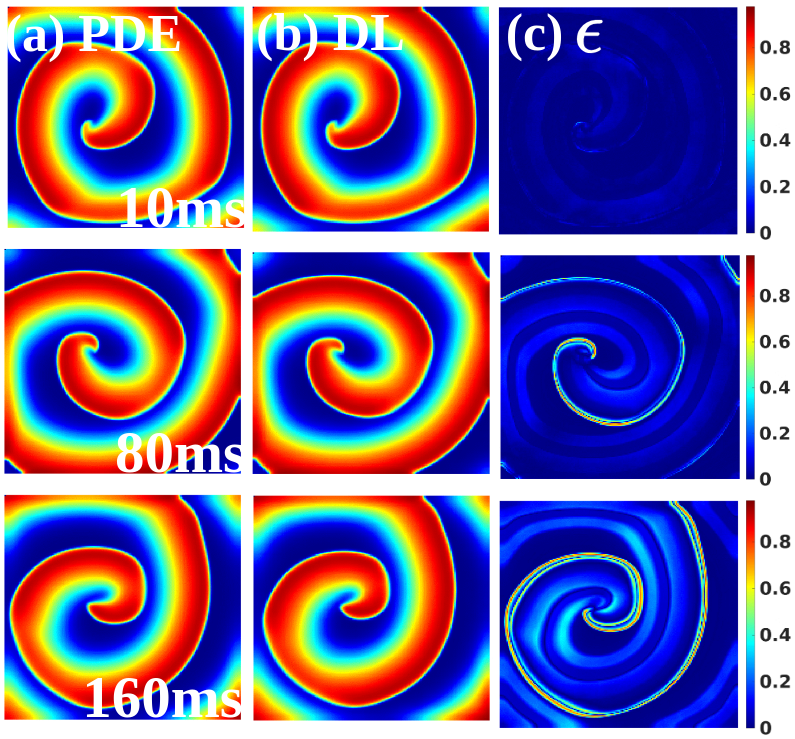}
\caption{\label{fig:fig2} Snapshots showing quasi-periodic spiral wave dynamics at different times (time stamps shown in the inset) in; (a) PDE, (b) DL model and (c) the difference $\epsilon$ between the PDE and DL. The difference in predictions, i.e., $\epsilon$ slowly increases over time.}
\end{figure}
We first predicted the evolution of the QP spiral wave for $\Delta t_{DL} = 10 ms$. A visual comparison is presented in Fig. \ref{fig:fig2}, where 
 the voltage is shown using a color scale with red/blue corresponding to activated/recovered tissue.
In this figure, the left column corresponds to the results from the PDE model, and thus the ground truth, and the middle column shows the DL predictions. The right column visualizes the difference between the PDE and DL snapshots and is a measure of the error $\epsilon$.
It shows that small errors were compounded over time. 

\begin{figure}
\includegraphics[scale=0.133]{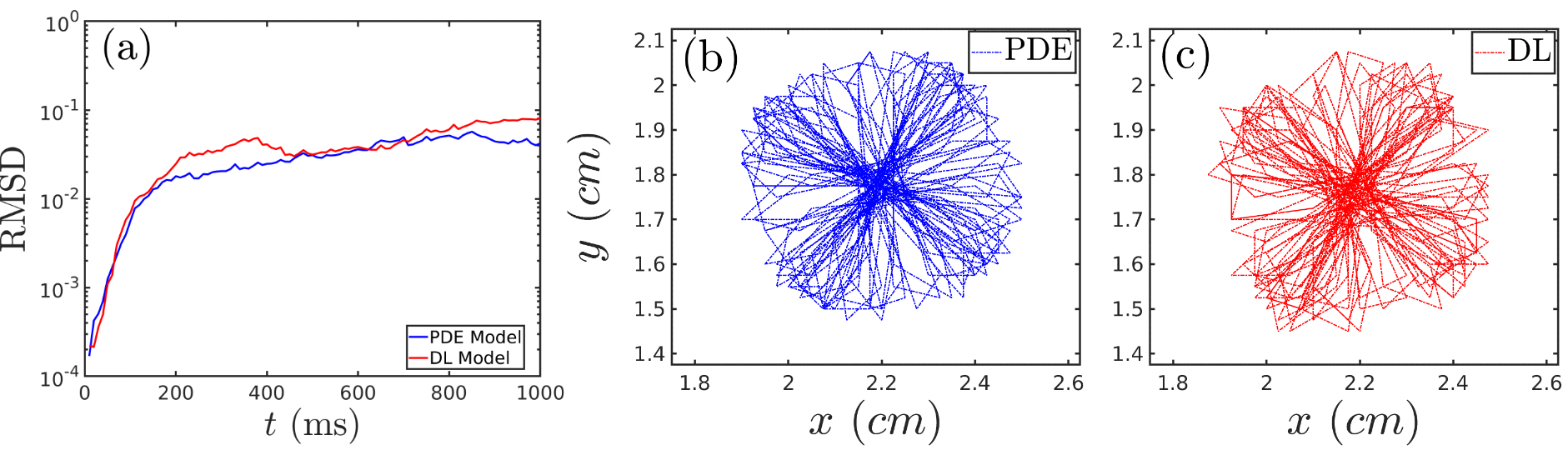}
\caption{\label{fig:fig3} (a) RMSD vs. time of a QP spiral wave obtained from a perturbed PDE 
simulation (blue curve) and from the DL model (red curve). (b) and (c): Tip trajectory of a QP spiral wave obtained using the PDE (b) and that predicted by the DL model (c). The  period corresponding to the dominant frequency of the time series in both the PDE and
the DL model is equal
and found to be $T_{D}=54 \ ms$.
}
\end{figure}

The increase of $\epsilon$ can be expected since  converting the initial $u$ field 
into a color-coded image, as required 
for the DL model, introduces 
small deviations from the ground truth.
We compared this DL error to the error from a PDE simulation in which the 
initial $u$ field is slightly perturbed. For this,  
we added random noise, uniformly distributed between 0 and 1 and with amplitude $10^{-5}$, to $u$ at each grid point, after which we continued to simulate Eq. 1. 
To quantify 
the rate at which the errors grow, we computed the root-mean-squared difference (RMSD)  using  
\begin{gather}
	\text{RMSD}(t) = \sqrt{\Large{\big{<}}\frac{1}{N^2}\sum_{i} \sum_{j} [u^{0}(i,j,t)-\Tilde{u}(i,j,t)]^{2}\Large{\big{>}}_{n}}
	\label{eqn:MSD}
\end{gather}
Here, $u^{0}$ represents the membrane potential of the unperturbed solution, and thus 
the ground truth, and  $\Tilde{u}$ represents the output of the DL model or the 
results of the perturbed PDE model. 
Thus, RMSD quantifies how rapidly the perturbations grow leading to changes in 
the wave dynamics. A comparison between the two errors is presented in 
Fig. \ref{fig:fig3}a, which shows that the error in the DL model grows at 
approximately the same rate as the error in the perturbed PDE model. 
In other words, the DL model is performing as well as the PDE model with added noise.

A further confirmation of the ability of the DL model to reproduce the 
PDE model is shown in Fig. \ref{fig:fig3}b and 3c. These panels show a comparison of the tip trajectory, which reveals that the flower pattern of the tip-trajectory of the quasi-periodic wave observed in the PDE model was accurately captured in our DL model. Furthermore, the time period corresponding to the dominant frequency of the time series of the spiral wave,  computed using the power spectrum, were the same in the two models ($54 ms$). 
Thus, the DL model can predict the quasi-periodic wave dynamics, its tip trajectory and its time period.  


\begin{figure}
\includegraphics[scale=0.295]{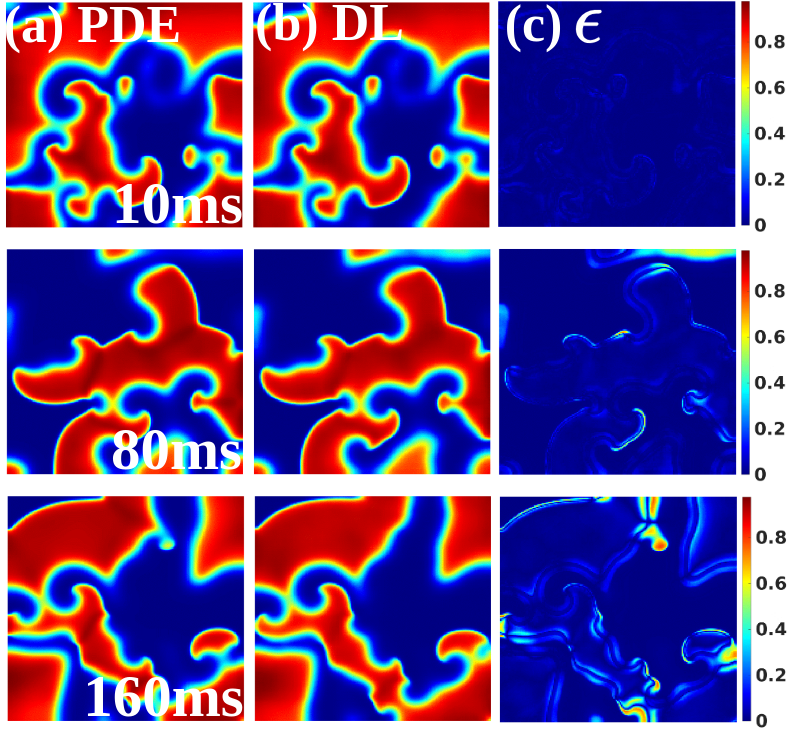}
\caption{\label{fig:fig4} Prediction of an SDC event using trained DL model in comparison with that from the PDE. One can observe the difference ($\epsilon$) increases with time.}
\end{figure}
We next tasked the trained network to predict  SDC wave dynamics.
In this case, we started with a snapshot taken from one of the 450 simulations
of the test set.  This snapshot was evolved in time using a
timestep $\Delta t_{DL}$ and we found that the wave activity in all simulations eventually terminated, even though the DL
model was not trained using termination snapshots (Fig. S2).  
Fig. \ref{fig:fig4} shows a comparison between the PDE and the DL model predictions and reveals that the DL model accurately predicts 
the activation pattern for short times, but start to deviate from the PDE model for longer times. 

\begin{figure*}
\includegraphics[scale=0.193]{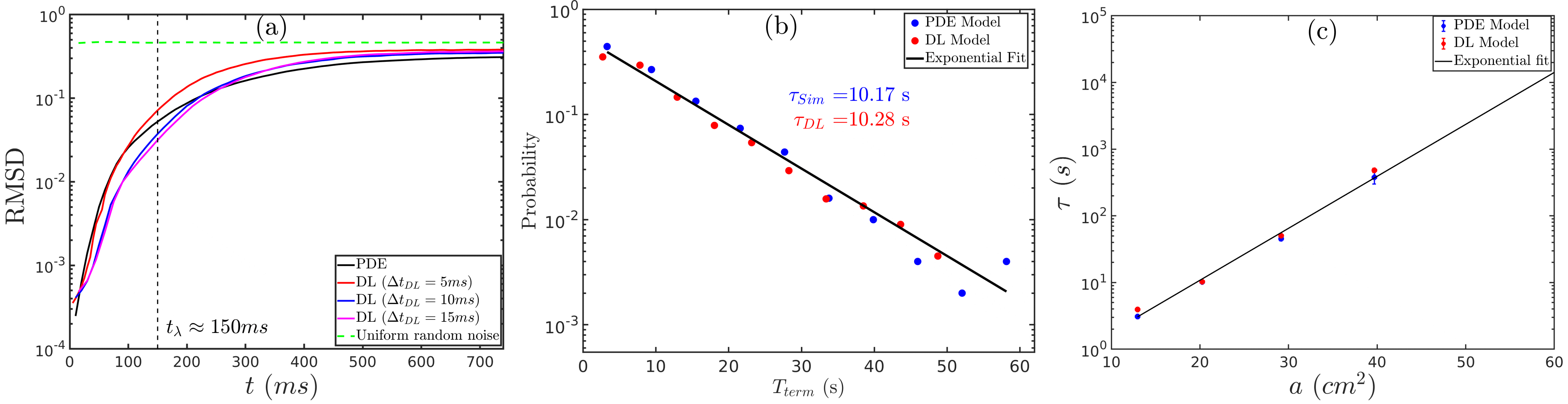}
\caption{\label{fig:fig5} (a): RMSD vs. time computed  for $n=100$ SDC events in both the PDE and the DL model. The Lyapunov time $t_{\lambda_{max}}$ is shown as a black dashed line. (b): Termination time statistics of the PDE model and the DL model ($\Delta t_{DL}= 10 ms$). The black line corresponds to an exponential fit. 
(c): Mean termination time for different domain sizes for $n=100$ events in both the PDE 
(blue symbols) and the  DL model (red symbols). The black line corresponds to an exponential fit.  }
\end{figure*}

These deviations can be expected and are hallmarks of SDC: small perturbations to an 
initial state grow exponentially fast, not only in the DL model but also in the 
underlying PDE model. To quantify 
the rate at which the perturbations grow, we computed the RMSD  in both the DL and PDE  model.   
For the latter, we again added 
 random noise, uniformly distributed between 0 and 1 and with amplitude $10^{-5}$, to $u$ at each initial grid point of the starting snapshot. 
In Fig. \ref{fig:fig5}a we plot the RMSD, averaged over 100 simulations, for the PDE model simulations (black line) and for the DL model for $\Delta t_{DL}=10 ms$ (blue line). 
Comparing these curves, we notice that both grow at approximately the same rate and approach the value 
obtained using randomly chosen values of $u^{pert}$  (green dashed line)   after a few rotations. 
This indicates that the propagation of perturbations in the $u$ variable observed in the PDE model is also captured by our DL model. 

The rate at which the noise-infused PDE and DL simulations deviate from the ground truth
can be compared to the Lyapunov exponent of the system. For this, we computed the largest Lyapunov exponent $\lambda_{max}$ from which we can determine 
 a characteristic  timescale $t_{\lambda}$~\cite{lilienkamp2017features,pathak2018model}. 
 This was computed by introducing a normalized perturbation to the variables $u$,  $v$, and $w$ in the PDE model and 
 monitoring its growth (\cite{skokos2009lyapunov}, see SM~\cite{SupMat_SDC}). The mean value of $\lambda_{max}$ was found to be approximately 6.6$s^{-1}$, 
 resulting in  $t_{\lambda} \approx 150 \ ms$ (Fig. S3). The value of $t_{\lambda}$ is shown as a vertical dashed line in Fig. \ref{fig:fig5}(a) which indicates that the DL model can predict reasonably well for times up to $t_{\lambda}$.
 
Earlier work has shown that termination in the PDE model is stochastic, with termination times 
that are exponentially distributed \cite{vidmar2019extinction}.  
Hence, we asked whether the termination time statistics of the DL model are comparable to those of the cardiac model. 
To address this questions, we evolved snapshots corresponding to different initial conditions in the DL model until
wave dynamics was terminated and recorded the termination time. 
Fig.\ref{fig:fig5}(b) shows the termination time distributions of SDC calculated using both the PDE and DL model. Both distributions follow an exponential distribution, indicated by black line, with a mean termination time of $\tau_{PDE}=10.2 s$ and $\tau_{DL}=10.3 s$, respectively. Furthermore, we have verified the time-averaged number of spiral wave tips 
in both the PDE and DL simulations was found to be nearly identical (7.7 vs 7.2). Thus, the DL model is able to 
faithfully capture the statistics of SDC. 

\begin{table}[t]
    \setlength{\tabcolsep}{2.2pt} 
    \renewcommand{\arraystretch}{1.05} 
    \begin{tabular}{||c|c|c|c||}
    \hline 
    { Dataset} & { $\tau_{fold} \ (s)$ } & { $\tau_{fold}  \ (s)$ } & { $\tau_{fold}  \ (s)$} \\[0.25ex]
    & ($\Delta t_{DL}= 5 ms$) & ($\Delta t_{DL}= 10 ms$) & ($\Delta t_{DL}= 15 ms$) \\
     \hline \hline 
       \textit{1}&  10.7 &  11.3 & 13.0\\
        \textit{2}& 9.3 & 11.4 & 31.0\\
        \textit{3}& 7.8 & 11.0 & 18.2\\
         \textit{4}& 10.8  & 9.3 & 23.9\\
          \textit{5}& 8.9 & 10.3 & 11.7\\ 
           \hline \hline 
       \textit{$\tau_{DL}  \ (s)$} &  9.5 &  10.7 & 19.6\\
     \hline
    \end{tabular}
    \caption{Mean termination time of each fold,  $\tau_{fold}$, and  averaged over the 5 folds of the DL model  trained using different values of $\Delta t_{DL}$. The ground truth corresponds to $\tau_{PDE}=10.2$s.}
         \label{Table:Tab1}
\end{table}

An important advantage of the DL model compared to the traditional PDE model is that the latter 
requires a small time step
due to the stiffness of the underlying equations while the former can perform predictions with a much larger time-step: $\Delta t_{DL}>> \Delta t_{PDE}$. 
To investigate the dependence of the DL model on this time step, we changed $\Delta t_{DL}$ and 
attempted to train the DL model. 
We found that the DL model  can be successfully trained  for $\Delta t_{DL} = 5, \ 10 \ \text{and}, \  15 \ ms$
using the same model architecture. 
However, training was challenging for $\Delta t_{DL} = 30 \ ms$ and the error did not decrease below 
$2\times 10^{-3}$, presumably due to reduced correlation between 
the input and output images.  
Figure.~\ref{fig:fig5}(a) shows the RMSD for different values of $\Delta t_{DL}$ (colored lines), which shows that the slope of the rate at which perturbations propagate is similar that of the PDE model (black line) for all values.
We also computed the termination time statistics for different $\Delta t_{DL}$ and the values of the mean termination time from the five different training folds, $\tau_{fold}$, are presented in Table.~\ref{Table:Tab1}. The DL model performed well for $\Delta t_{DL}=5 \ \text{and} \ 10 \ ms$ but started to deviate from the ground truth for  $\Delta t_{DL}=15 ms$. The above results indicate that the DL model can capture the statistics of SDC as long as  $\Delta t_{DL}$ is 
not too large.

Finally, we asked whether the DL model trained on one  domain size can be used to predict
SDC statistics on another domain size. 
As for the original domain size, we found that 
the dynamics on a domain that was 
4 times larger could be predicted for approximately 
one Lyapunov time while the RMSD vs. time 
for the DL model is similar to the one obtained
from the PDE model (Figs. S4 and S5).
We also
 computed the mean termination time from PDE simulations on square $N \times N$ domains with 
$N=144,216,252,324$, and $360$. Consistent with earlier work \cite{qu2006critical,vidmar2019extinction}, 
this mean termination time increases exponentially with the domain size $a=L^2$ (blue symbols, Fig.~\ref{fig:fig5}(c)).  
Using the snapshots from these PDE simulations, we then computed the mean termination time for the different 
domain sizes using the trained DL model. 
The results, shown as red symbols in Fig.~\ref{fig:fig5}(c), demonstrate that the DL model also captures this 
exponential increase and is consistent with the PDE model.

Our DL model only uses images of a single variable and is therefore agnostic to the 
details of the electrophysiological model used to produce these images.  
This means that it should perform equally well for more complex and detailed models, for which the 
computational gains may become considerable. 
To illustrate this,  we compared the time required to simulate  1 s of  wave dynamics in three electrophysiological models of increasing complexity and to the required time in our DL model (Fig. S6 in SM~\cite{SupMat_SDC}). 
This comparison revealed that the DL model can simulate wave dynamics in near real-time, and that the 
gain in computational time in a detailed model can be several orders of magnitude, 
especially for large domains.


In summary, we have shown that a single DL model can learn both  QP and SDC wave dynamics and can be 
used to predict patterns using 
a much larger time step than required to simulate the underlying PDEs, potentially 
saving valuable computational time. 
We also showed that the DL model was able to accurately predict the tip trajectory of the QP spiral wave and can 
simulate termination, even though the training set 
did not contain termination events. Additionally, the statistics of the SDC computed using the DL model is 
identical to the one from the PDE model. 
Our approach does not depend on the details of the model used to generate training data and should, 
therefore, be applicable to a wide range of models. Furthermore, 
the model was trained using data from a single variable, even though the 
PDE model contains more variables. 
This could make
our approach to be useful when applied to experimental and clinical data, which 
typically only records a single variable. 

\vspace{0.25cm}
This work was supported by National Institutes of Health Grant R01 HL122384 and RO1	HL149134.

%

\end{document}



\title{ \ \hspace{1.95cm} Supplemental Material \newline Prediction of excitable wave dynamics using machine learning
}

\author{Mahesh Kumar Mulimani ,Sebastian Echeverria-Alar,Michael Reiss ,
Wouter-Jan Rappel}
 \affiliation{ Department of Physics, UC San Diego, La Jolla, CA, United States } 

\date{\today}

\maketitle



\subsection{\label{sec:FK_Model} Fenton-Karma Model}
We used the Fenton-Karma (FK) model, consisting of the
three variables $u, v$ and $w$ (for details see Ref.~\cite{fenton1998vortex}). The model equations are written as:

  \begin{align*}
  \label{eqn:u}
  \frac{\partial{u}}{\partial{t}} &  = D_0 {\nabla^2}u - [J_{fi}(u,v) + J_{so}(u) + J_{si}(u,w)]  \numberthis \\[1.9pt] 
  \label{eqn:v}
   \frac{\partial v}{\partial{t}} & = \frac{\Theta(u_c-u) (1-v)}{\tau_{v}^{-}(u)} - \frac{\Theta(u-u_c) \ v}{\tau^{+}_{v}}  \numberthis \\ 
   \label{eqn:w}
    \frac{\partial w}{\partial{t}} & = \frac{\Theta(u_c-u)(1-w)}{\tau^{-}_{w}} - \frac{\Theta(u-u_c) w}{\tau^{+}_{w}} \numberthis \\ \\
  J_{fi}(u,v) & = -\frac{v}{\tau_d} \Theta(u-u_c) (1-u) (u-u_c) \nonumber \\
   J_{so}(u) & = -\frac{u}{\tau_o} \Theta(u_c-u) + \frac{1}{\tau_r} \Theta(u-u_c) \nonumber \\
   J_{si}(u,w) & = -\frac{w}{2 \tau_{si}}(1 + tanh[k(u-u_c^{si})]) \nonumber \\
   \tau_{v}^{-}(u) & = \Theta (u_v-u) \ \tau_{-v2} + \Theta (u-u_v) \ \tau_{-v1}\\
\Theta(u_c-u) = \begin{cases}
        1 & \text{if } u \geq u_c\\
        0 & \text{if } u < u_c 
    \end{cases} \\
   \quad  \Theta(u_v-u) = 
   \begin{cases}
        1 & \text{if } u \geq u_v\\
        0 & \text{if } u < u_v 
    \end{cases}    
   \label{eqn:pde}
  \end{align*}

These equations were simulated using two different parameter sets. One set 
results in a quasi-period  (QP) spiral wave  while 
the other one results in spiral defect chaos (SDC) (see also the main text).
These parameters are given in Table~\ref{tab:table_parameters}.
\begin{table}[]
    \setlength{\tabcolsep}{10pt} 
    \renewcommand{\arraystretch}{1.05} 
    \begin{tabular}{||c|c|c||}
    \hline 
        {\B Parameter name} & {\B Parameter value (SDC)} & {\B Parameter value (QP)}\\ [0.25ex] 
     \hline \hline
        \large$u_m$ & \La 1 & \La 1\\ 
       \large$\tau_v^{+}$ &  \La3.3 &  \La3.3\\
        \large$\tau_{-v1}$ & \La12 &  \La12\\ 
        \large$\tau_{-v2}$ & \La2 &  \La2\\   
        \large$\tau_{+w}$ & \La1000 &  \La1000\\ 
        \large$\tau_{-w}$ & \La100 &  \La500\\ 
        \large$\tau_{d}$ & \La0.36 &  \La0.3\\ 
        \large$\tau_{0}$ &  \La5 &  \La5\\
        \large$\tau_{r}$ & \La33 &  \La33\\ 
        \large$\tau_{si}$ & \La29 &  \La29\\   
        \large$\kappa$ & \La15 &  \La15\\ 
        \large$u_{c}^{si}$ & \La0.7 &  \La0.7\\ 
        \large$u_{c}$ & \La0.13 &  \La0.11\\    
        \large$u_{v}$ & \La0.04 &  \La0.04\\         
     \hline
    \end{tabular}
   \quad
    \setlength{\tabcolsep}{10pt} 
    \renewcommand{\arraystretch}{1.05} 
    \caption{ FK-model parameters for SDC and QP spiral wave simulations.}
         \label{tab:table_parameters}
\end{table}

\subsection{DL Model}

\subsubsection{2D convolutional encoder-decoder network}
We used a 2D convolutional encoder-decoder neural network model for learning the spatiotemporal dynamics, schematically shown in Fig. 1 of the main text.
The input for our DL model is an image of the $u$ variable for both a SDC simulation and a QP  spiral wave. The 
DL model attempts to predict the image corresponding to the $u$ variable 
at a later time $t+\Delta t_{DL}$. Note that $\Delta t_{DL}$ can be 
taken to be much larger that the time step required to solve the FK model:
$\Delta t_{DL}>>\Delta t_{PDE}$

The encoder network comprised of four sets of stacked 2D convolutional (conv2D) and max-pool layers with a batch normalization layer at the end of the network. These sets of layers reduce the dimension of an input image successively using filters while retaining the essential information.
The output  from this encoder network  was passed to a decoder network, which  
consisted of four sets of 2D convolutional transpose and 2D convolutional layers stacked together with a final 3D Convolutional layer. The 2D convolutional transpose layer increases the dimension of the  input from the previous layers using filters (functionally opposite to the max-pool layer in the encoder network) thereby restoring the dimension at the end of the decoder network. 
Thus, the encoder-decoder network learns the mapping between the input and the output image which are time-shifted.

\subsubsection{\label{sec:DL_Model} Training details of DL-Model}  
Training of the DL model was considered to be successful if the target wave pattern was predicted by the DL model through fitting the values of $DL_{Param}$ parameters. If the expected target wave pattern was not achieved, the $DL_{Param}$ values were changed by optimizers such as Adam (Adaptive momentum estimation) ~\cite{kingma2014adam}. The difference between the expected outcome and the prediction was calculated using a loss function. In our DL model, we used a Huber loss function that was a mixture of mean squared error (MSE) and mean average error (MAE) ~\cite{chollet2015keras}. 
We list some of the hyperparameter values used in training our DL model in Table.~\ref{tab:table_DL_parameters}.
The dataset for training the DL model was segregated into training  and validation data. The training data is the data on which our DL model learned and for which the parameters $DL_{Param}$ were fit. The validation data, which is not used for fitting the $DL_{Param}$, was used to check if the DL model had actually learned.

\begin{figure*}
\includegraphics[scale=0.275]{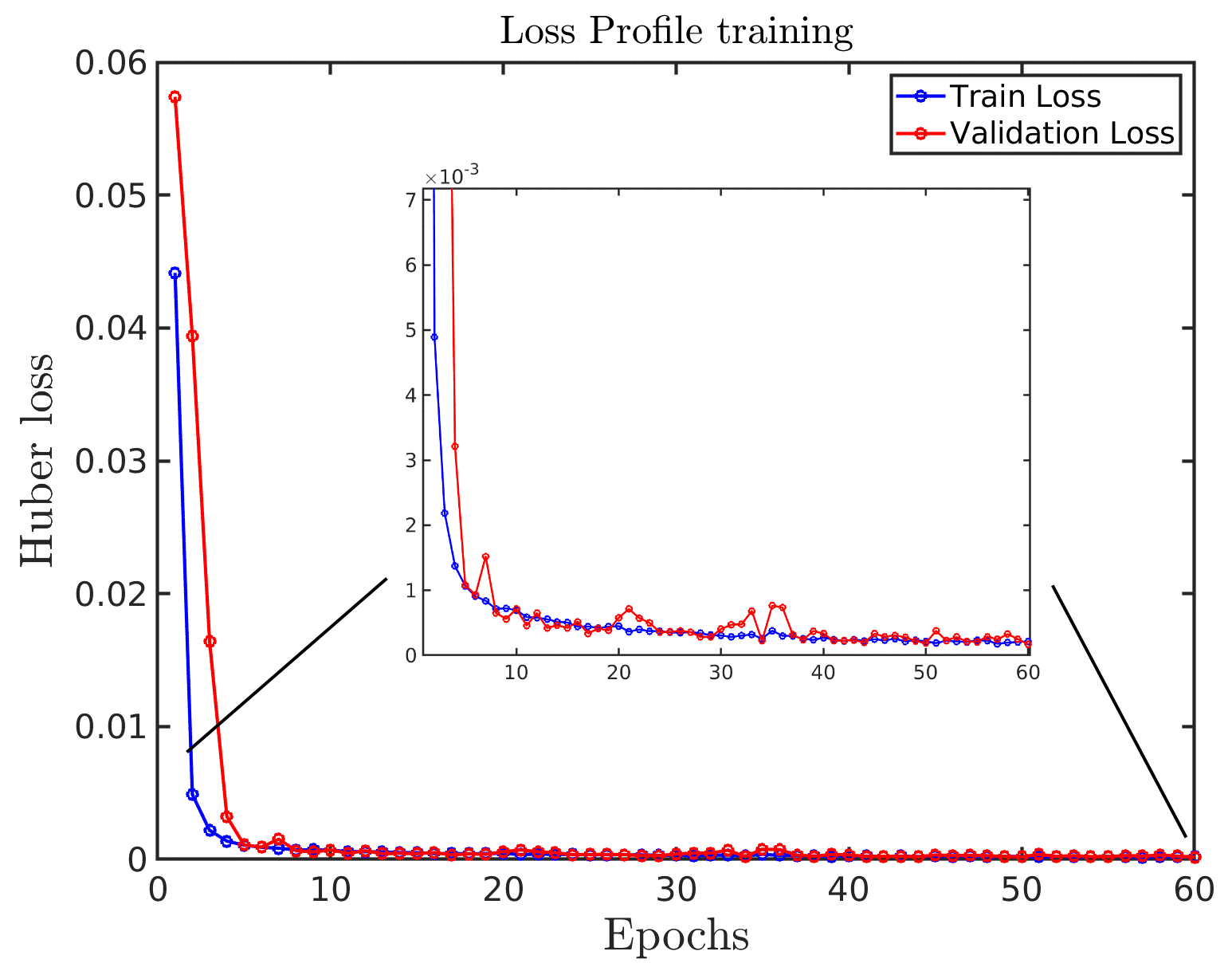}
\caption{\label{fig:loss} The loss profile of a DL model training. In the inset figure, we show the slow decrease of both the losses after 10 epochs that ultimately converged.} 
\end{figure*}

If the training loss reached a value of $ \leq 2 \times 10^{-4}$ and did not furthermore decrease with 10 epochs, we concluded that the training process is completed. 
We show a typical loss profile of the training process in Fig.~\ref{fig:loss}. 

\begin{table}[]
    \setlength{\tabcolsep}{10pt} 
    \renewcommand{\arraystretch}{1.05} 
    \begin{tabular}{||c|c||}
    \hline 
        {\B Hyperparameter name} & {\B Hyperparameter value} \\ [0.25ex] 
     \hline \hline
       \large\textit{learning rate} &  0.001 \\
        \large\textit{batch size} & 5 \\
        \large\textit{Epochs} & 60 \\
    
     \hline
    \end{tabular}
    \caption{Table of hyperparameters that are used in training our DL model.}
         \label{tab:table_DL_parameters}
\end{table}

\subsubsection{Datasets}
As mentioned in the main text, we chose only a single variable in our training
data, the trans-membrane potential $u$, and asked  whether using this scalar spatiotemporal variable is sufficient to train a DL model. 
In other words, the DL model does not consider additional variables, including ones
that represent ionic currents, and should therefore be applicable to a wide
variety of models that display QP spiral waves and SDC. 

\begin{figure*}
\includegraphics[scale=0.220]{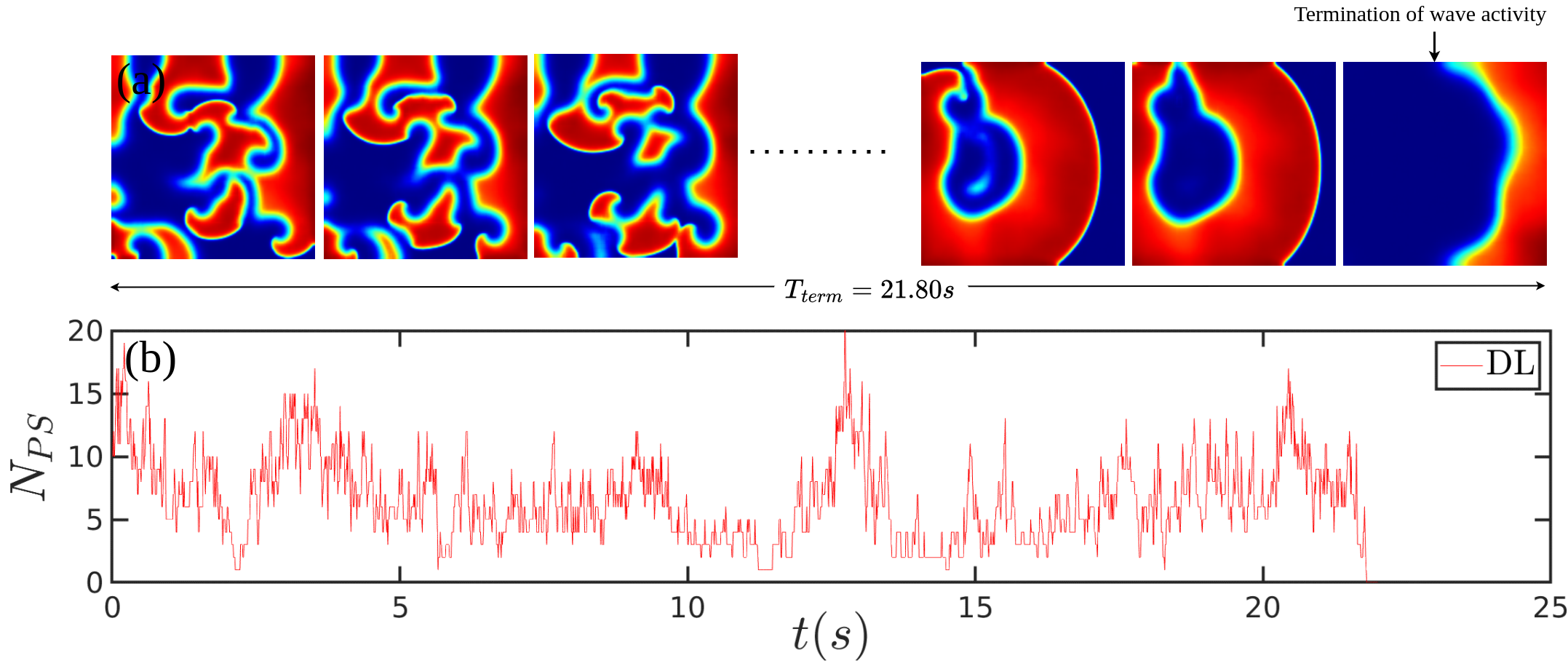}
\caption{\label{fig:figS5} (a) Prediction sequence of SDC wave dynamics from the DL model exhibiting termination (occurring at $t=21.8 \ s$). (b) Corresponding number of phase singularities as a function of time.}
\end{figure*}

Our DL model was trained for two different patterns: a quasi-periodic spiral wave  and chaotic wave dynamics marked by spiral defect chaos.  
In the latter training set, we did not include any termination event. Nevertheless, the DL model was able to 
``discover" termination, indicating that it has learned the excitable dynamics of the model. This is illustrated in Fig. \ref{fig:figS5}, which shows an example of a termination sequence predicted by the DL model. 
For the training set, we only used the first  400 snapshots from each SDC event with duration
400$\Delta t_{DL}$, where  $\Delta t_{DL}$ is the timestep in the DL model. 
In this set, we only considered SDC events that were at least 1 s longer than this duration.   
 
 The $u$ spatial snapshot data was represented as an unsigned integer 8 bit (UINT8) RGB image (as shown in Fig.1 of the main article). Using RGB images was motivated twofold. First, spatiotemporal data can be easily trained  using DL models if the data format is represented as images instead of the raw data. Second, experimental and 
 clinical data are typically represented by images that record a single variable.

\subsection{\label{sec:FK_Model} Calculation of the maximum finite-time Lyapunov exponent}

The spatiotemporal dynamics exhibited by the solution $\{\bar{u},\bar{v},\bar{w}\}$  of Eqs.~(\ref{eqn:u}-\ref{eqn:w})
can be characterized with the maximum Lyapunov exponent $\lambda_{1}$. However, this dynamical quantity can only be extracted if the Heaviside functions $\Theta(u-u_{m})$, where $m=\{c,v\}$, in Eqs.~(\ref{eqn:u}-\ref{eqn:w}) are approximated by the analytical expression $0.5(1 + tanh(s_{f}(u-u_{i})))$. This approximation is equivalent to the Heaviside function when the stiffness parameter $s_{f}$ tends to infinity. Numerically, we  have verified that using the value $s_{f}=100$ is a reasonable approximation by comparing the distributions of termination times. 
\begin{figure}
\includegraphics[scale=0.325]{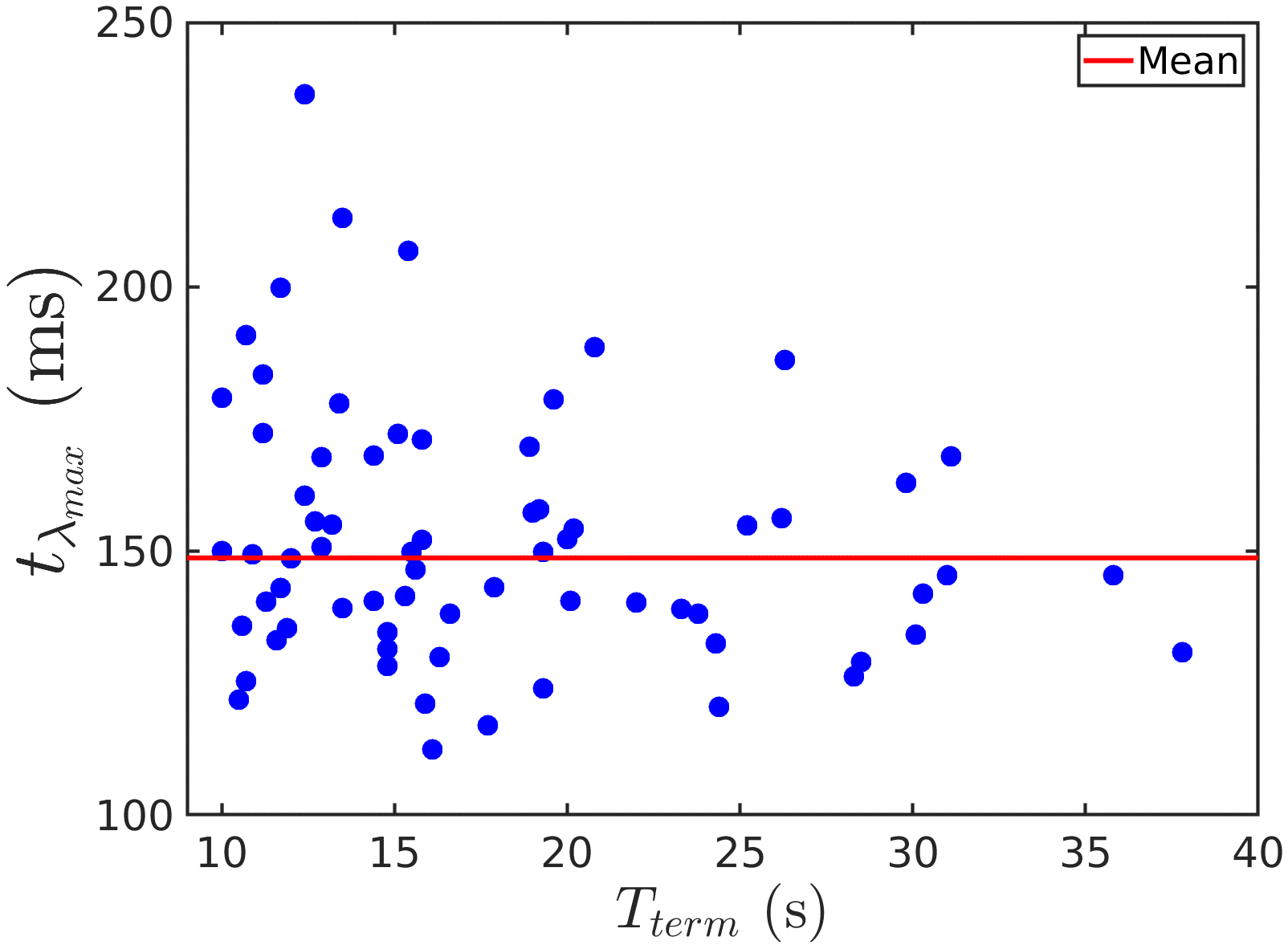}
\caption{\label{fig:Lyapunov} The Lyapunov time values calculated for 100 different SDC events. The mean value is shown as a red line.}
\end{figure}


The maximum Lyapunov exponent was obtained by following the numerical procedure detailed in~\cite{skokos2009lyapunov}. A normalized perturbation around $\{\bar{u},\bar{v},\bar{w}\}$ was performed and then monitored on time. The perturbation is governed by the linearized equation $\partial_{t}\delta\mathbf{\{\bar{u},\bar{v},\bar{w}\}}=\mathbf{J}(\{\bar{u},\bar{v},\bar{w}\})\delta\mathbf{\{\bar{u},\bar{v},\bar{w}\}}$, 
where $\delta\mathbf{\{\bar{u},\bar{v},\bar{w}\}}$ is a perturbation around the solution $\{\bar{u},\bar{v},\bar{w}\}$ and $\mathbf{J}(\{\bar{u},\bar{v},\bar{w}\})$ is the respective discretized Jacobian. 
Every $\Delta\tau \sim 100$ ms, an orthonormalization of the perturbation was carried out. After repeating the previous algorithm for $k$ iterations, the Lyapunov exponent can be approximated by 

\begin{equation}
\lambda_{i}=\frac{1}{k\Delta\tau}\sum_{i=1}^{k}\ln[|\delta\mathbf{\{\bar{u},\bar{v},\bar{w}\}}|],
\label{m1}
 \end{equation}
when $k$ is sufficiently large. In our case, $k\Delta\tau$ was bounded by the termination time. Fig. \ref{fig:Lyapunov} shows the results for 100 SDC events.


\subsection{\label{sec:Termination_SUpMat} DL model applied to larger domain size }
\begin{figure*}
\includegraphics[scale=0.383]{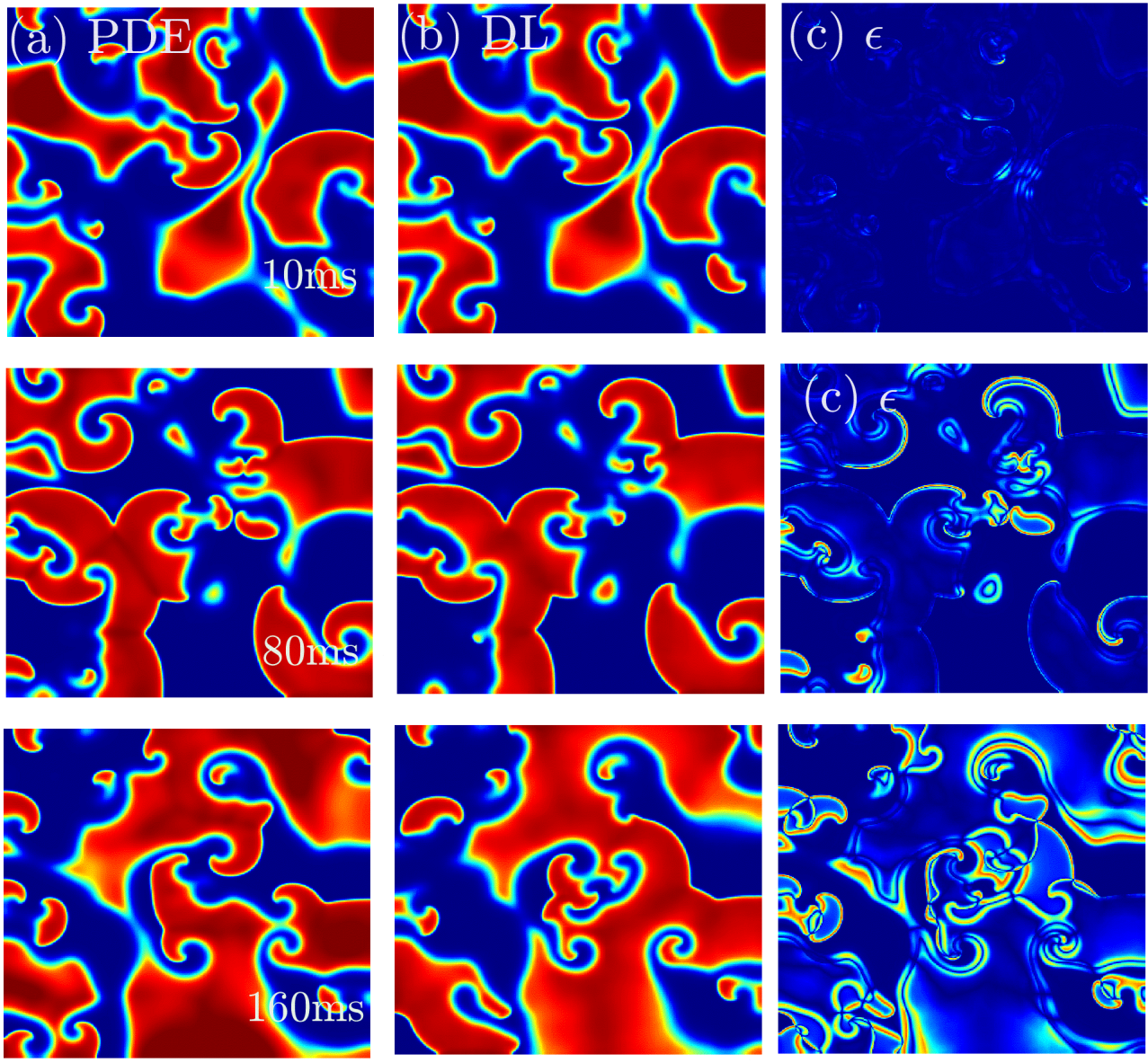}
\caption{\label{fig:figS6} Prediction of SDC wave dynamics for a bigger domain size ($N^{\prime}=360$) using the DL model trained on $N=180$.}
\end{figure*}

\begin{figure*}
\includegraphics[scale=0.383]{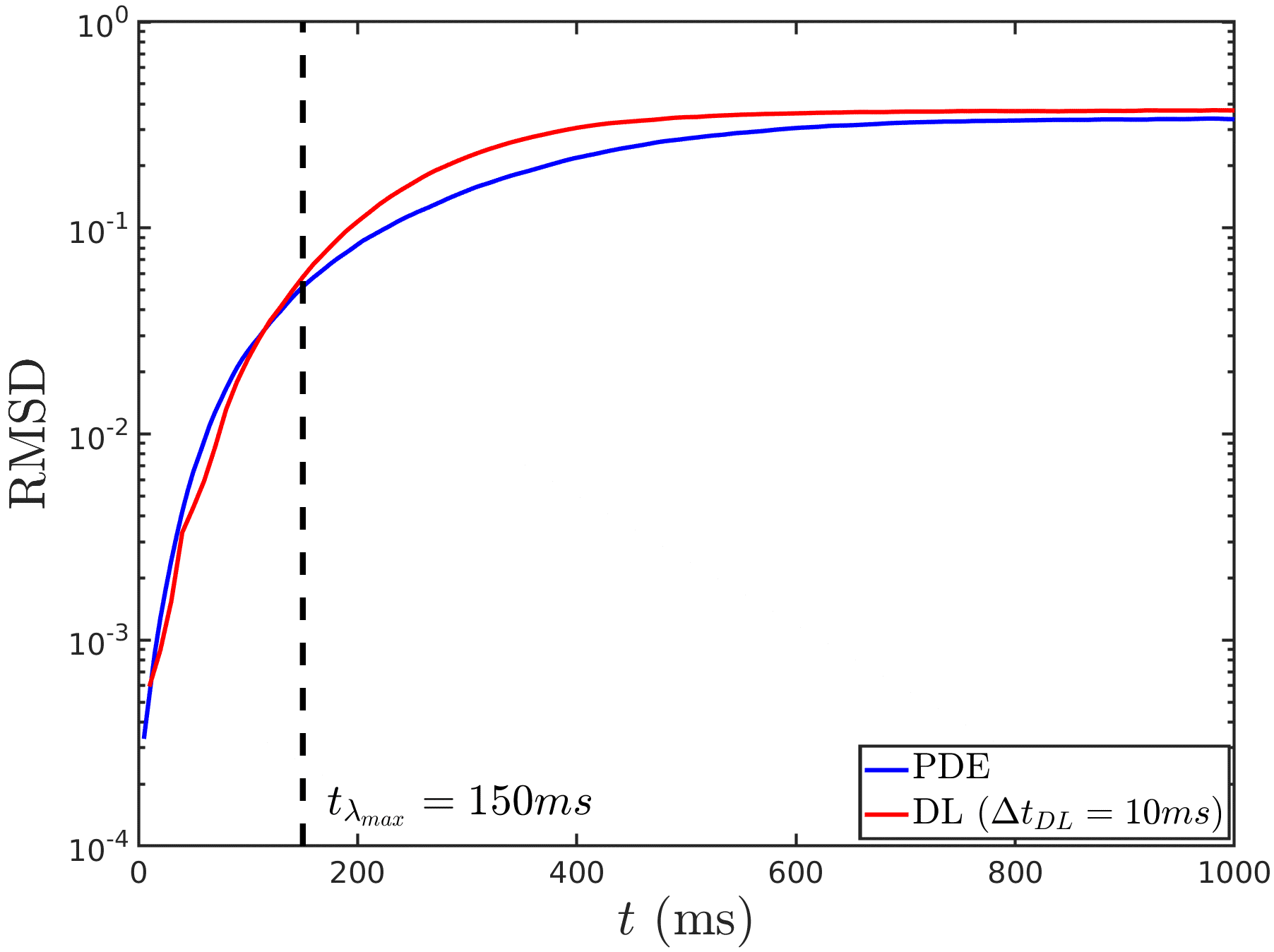}
\caption{\label{fig:figS7} RMSD of SDC  (averaged over $n=100$ events) for
the  DL model (domain size $N^{\prime}=360$)  and for a perturbed PDE  model 
using a domain size of $N^{\prime}=360$.
 }
\end{figure*}

We have verified that the application of the trained DL model is not restricted to the domain size on which it was trained (i.e., 
$N \times N$ with $N = 180$).
In fact, we can use the trained DL model to predict  termination time statistics for different domain sizes $N^{\prime}$, creating a ``scaled" DL model.
In Fig.~\ref{fig:figS6} we show the predictions for $N^{\prime}=360$ using the DL model that is trained on $N = 180$. 
As in the case of $N=180$, the DL model is able to predict the dynamics up to approximately the maximum 
Lyapunov time (cf. Fig. 5a in the main text). 
Furthermore, we calculated the RMSD (see Eq. 2 in the main text) for $N^{\prime}=360$, to verify whether our scaled DL model captures the rate of propagation of the perturbations in the $u$ variable
observed in the PDE model. The results are shown in Fig.~\ref{fig:figS7}A, which 
demonstrates that rate of propagation of  perturbations in the DL and PDE model are 
similar.

\subsection{\label{sec:Comparison_CUDA_vs_Sim} Time comparison of the DL model with cardiac electrophysiological models }

\begin{figure*}
\includegraphics[scale=0.42]{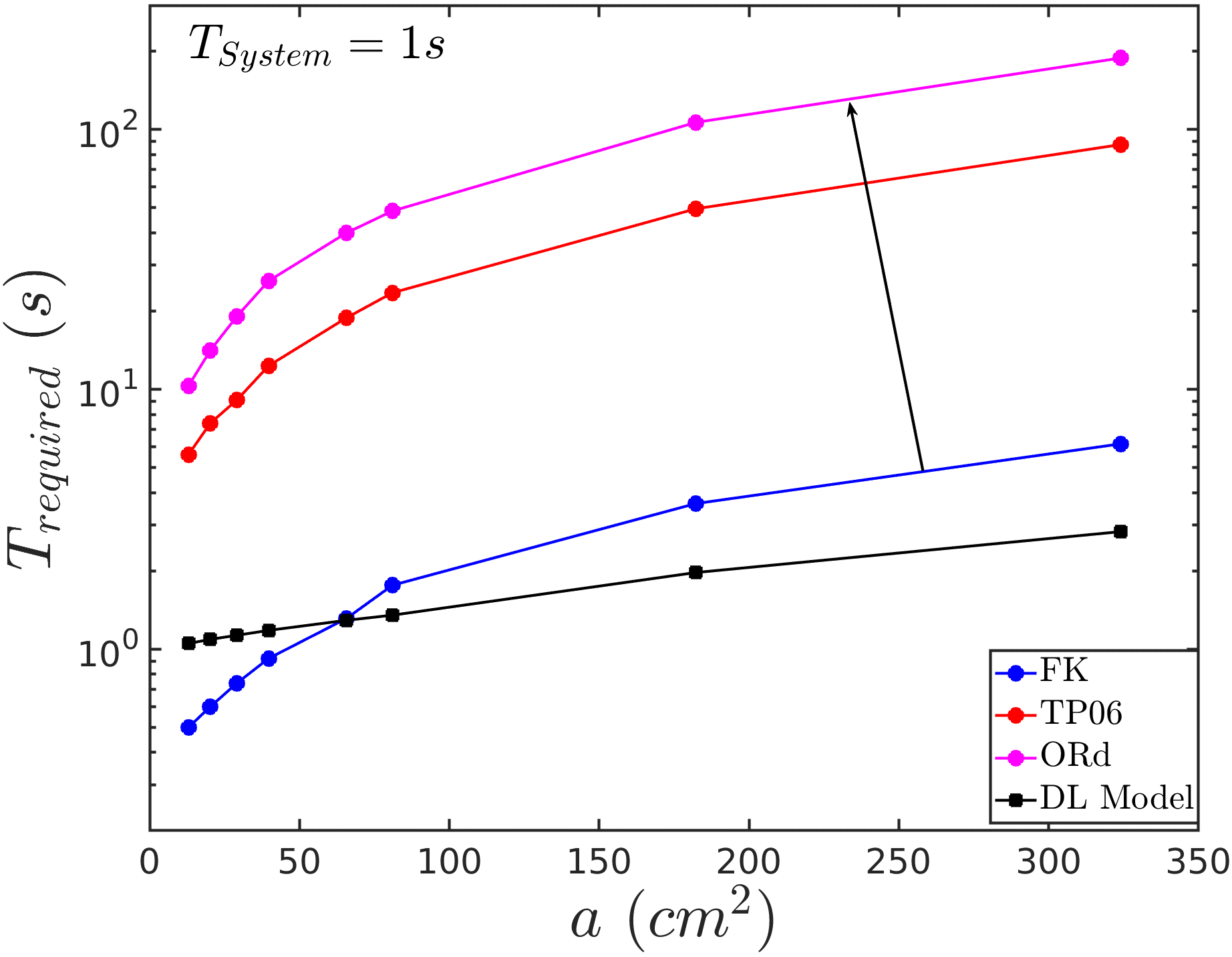}
\caption{\label{fig:figs8} Computation time of different electrophysiological models and 
the DL model as a function of domain size. The arrow indicates the direction of increasing model complexity.}
\end{figure*}

Cardiac electrophysiological models are becoming increasingly complex. For example, the ten Tusscher and Panfilov 
(TP06) model   published in 2006 had 17 state variables and 53 parameters, the O'Hara-Rudy (ORd) model published in 2011 has 26 state variables and $ > 50$ parameters, and the Asakura model published in 2014 has 45
state variables and  $\approx 100$ parameters. Clearly, 
simulating electrical activity in two- and three-dimensional tissue using these models 
is more intensive than using simplified models. 
In Fig.~\ref{fig:figs8} we compare  the  time required  to perform a 2D simulation 
for 1 s of these  electrophysiological models, using parallelized algorithms,  to simulation time required for the DL model.
We  observe that for large domain sizes the DL model shows an  advantage in computational time over all electrophysiological 
models. This advantage becomes more pronounced for the detailed models, where it can approximate 
two orders of magnitude for the largest domain size considered here. 

%